\documentclass[12pt,final]{article}
\usepackage{graphicx}
\usepackage{epsfig}
\usepackage{dcolumn}
\usepackage{bm}
\usepackage{achemso}
\usepackage{amssymb,amsmath}
\usepackage{gensymb}
\usepackage{multirow}
\usepackage{booktabs}
\makeatletter
\def\@biblabel#1{(#1)}
\makeatother
\usepackage{setspace}

\usepackage{booktabs,rotating}
\usepackage{fancyhdr}
\usepackage{booktabs,caption}
\usepackage[flushleft]{threeparttable}
\usepackage{enumerate,subfigure,tabularx,longtable}
\usepackage{longtable}
\usepackage[utf8x]{inputenc}
\UseRawInputEncoding
\usepackage{gensymb}
\usepackage {threeparttable} 

\usepackage{natbib}
\setkeys{acs}{articletitle = true}

\newcommand{\comment}[1]{}

\def\gsim {\mbox{\hbox{ \lower-.6ex\hbox{$>$}
\kern-1.12em \lower.5ex\hbox{$\sim$}\kern+.35em}}}
\def\lsim {\mbox{\hbox{ \lower-.6ex\hbox{$<$}
\kern-1.12em \lower.5ex\hbox{$\sim$}\kern+.35em}}}
\makeatletter
\let\@fnsymbol\@arabic
\makeatother

\begin{document}


\title{\vspace{-3.0cm} Molecular modeling of interfacial properties \\
of the hydrogen+water+decane mixture  \\
in three-phase equilibrium
}

\author{Yafan Yang\thanks{Corresponding Author, email address: yafan.yang@cumt.edu.cn} $^{,\dag}$,
Jingyu Wan\thanks{Corresponding Author, email address: jingyu\_wan123@163.com} $^{,\dag}$,  
Jingfa Li$^\ddag$,\\
Guangsi Zhao$^\dag$,
and Xiangyu Shang\thanks{Corresponding Author, email address: xyshang@cumt.edu.cn} $^{,\dag}$\\
$^\dag$ State Key Laboratory for Geomechanics and Deep \\
Underground Engineering, China University of Mining \\
and Technology, Xuzhou 221116, China. \\
$^\ddag$ School of Mechanical Engineering and Hydrogen\\
Energy Research Center, Beijing Institute of \\
Petrochemical Technology, Beijing 102617, China. \\
}

\date{\today}
\maketitle
\newpage
\begin{abstract}

The understanding of geochemical interactions between H$_2$ and geofluids is of great importance for underground H$_2$ storage, but requires further study. We report the first investigation on the three-phase fluid mixture containing H$_2$, H$_2$O, and n-C$_{10}$H$_{22}$.
Molecular dynamics simulation and PC-SAFT density gradient theory are employed to estimate the interfacial properties under various conditions (temperature ranges from 298 to 373 K and pressure is up to around 100 MPa).
Our results demonstrate that interfacial tensions (IFTs) of the H$_2$-H$_2$O interface in the H$_2$+H$_2$O+C$_{10}$H$_{22}$ three-phase mixture are smaller than IFTs in the H$_2$+H$_2$O two-phase mixture.
This decrement of IFT can be attributed to C$_{10}$H$_{22}$ adsorption in the interface.
Importantly, H$_2$ accumulates in the H$_2$O-C$_{10}$H$_{22}$ interface in the three-phase systems, which leads to weaker increments of IFT with increasing pressure compared to IFTs in the water+C$_{10}$H$_{22}$ two-phase mixture.
In addition, the IFTs of the H$_2$-C$_{10}$H$_{22}$ interface are hardly influenced by H$_2$O due to the limited amount of H$_2$O dissolved in bulk phases.
Nevertheless, relatively strong enrichments and positive surface excesses of H$_2$O are seen in the H$_2$-C$_{10}$H$_{22}$ interfacial region.
Furthermore, the values of the spreading coefficient are mostly negative revealing the presence of the three-phase contact for the H$_2$+H$_2$O+C$_{10}$H$_{22}$ mixture under studied conditions.

\end{abstract}

KEYWORDS: Three phase mixture; Interfacial tension; Square gradient theory; PC-SAFT; Molecular simulation; Underground H$_2$ storage.\\

\newpage
\section{Introduction}

The discharge of greenhouse gases into the air through the burning of fossil fuels is a major environmental issue since it contributes to global warming.\cite{houghton2005global,vicente2020review}
The progressive substitution of fossil fuels with hydrogen as a clean fuel has garnered significant attention.
Utilizing renewable resources for hydrogen production has the potential to significantly decrease carbon emissions, thereby playing an important role in mitigating climate change.\cite{momirlan2002current,midilli2005hydrogen,caglayan2020technical}

A significant challenge lies in the limited hydrogen storage capacity in times of excess production.\cite{tarkowski2019underground,zivar2021underground,sambo2022review}
Over the years, there has been successful implementation of industrial-scale storage of natural gas in geological locations, such as in depleted gas/oil reservoirs (80\%), saline aquifers (12\%), and salt caverns (8\%).\cite{thomas2015carbon}
Nonetheless, when it comes to storing pure H$_2$ at an industrial scale, only salt caverns have been employed so far.\cite{lankof2020assessment} There have been limited instances where other sites like aquifers and depleted gas reservoirs have been used effectively to store gas mixtures containing H$_2$.\cite{panfilov2006underground,pichler2019underground}
Therefore, exploring alternative storage site options holds immense importance and warrants further investigation.

Depleted oil reservoirs have proven successful in storing natural gas and CO$_2$,\cite{thomas2015carbon,tarkowski2021storage} but they have not been utilized for H$_2$ storage. Unlike CH$_4$ and CO$_2$, H$_2$ exhibits large minimum miscibility pressures with oil.\cite{florusse2003solubility,yellig1980determination,rao2003determination} Consequently, the geological storage of H$_2$ in depleted oil fields is mostly expected to encounter a three-phase fluid system.
The interfacial characteristics of multiphase fluid systems play a critical role in determining the capillary entry pressure required for gas penetration in the pores.\cite{hosseini2022capillary,yekta2018determination}
This, in turn, significantly affects the efficiency of the capillary sealing of the caprocks in depleted oil reservoirs, making it one of the key influencing parameters.\cite{hosseini2022capillary,pan2021underground,lubon2021influence}

Recently, many investigations, both experimental\cite{chow2018interfacial,chow2020erratum,hosseini2022h2,slowinski1957effect,massoudi1974effect,isfehani2023interfacial,muhammed2023hydrogen,mirchi2022interfacial} and simulation-based\cite{van2023interfacial,yang2023molecular,doan2023molecular}, have been conducted to examine the interfacial characteristics in two-phase mixtures that involve both hydrogen and water.
The interfacial tension (IFT) of the hydrogen+water mixture exhibits a nearly linear reduction in response to increments of pressure and temperature, and at elevated temperatures, the decrease in IFT from rising pressure is less pronounced.\cite{chow2018interfacial,chow2020erratum,hosseini2022h2,slowinski1957effect,massoudi1974effect}
The IFTs of H$_2$+brine systems show a proportional increase as the salinity rises.\cite{hosseini2022h2,isfehani2023interfacial}
Moreover, the inclusion of impurities such as N$_2$, CO$_2$, and/or CH$_4$ in the gas+H$_2$+H$_2$O two-phase systems leads to decrements in IFTs.\cite{muhammed2023hydrogen,chow2018interfacial,mirchi2022interfacial,isfehani2023interfacial}
In addition, molecular dynamics (MD) simulations predict IFTs in the hydrogen+water system with reasonable accuracy comparing experimental data.\cite{van2023interfacial,yang2023molecular,doan2023molecular}
Importantly, adsorption of hydrogen in the interfacial region of the hydrogen+water mixture has been observed based on the density distributions from MD simulations, which explains the relatively small IFT in contrast to the surface tension of pure H$_2$O.\cite{yang2023molecular}

Meanwhile, several studies have focused on the characteristics of interfaces in the gas+H$_2$O+oil three-phase mixtures.\cite{bahramian2007vapour,pereira2016interfacial,pereira2014vapor,amin1998interfacial,pan2023experimental,yang2005interfacial,yang2022studyJML2,yang2022interfacialatmosphere,lobanova2016saft,yang2023interfacialJML}
Three different interfaces are involved in the gas+H$_2$O+oil three-phase mixtures, namely, the gas-H$_2$O, the gas-oil, and the H$_2$O-oil interfaces.
Generally, the IFTs of the CH$_4$-H$_2$O interface in three-phase systems containing CH$_4$/CO$_2$+H$_2$O+oil\cite{yang2022studyJML2,yang2023molecular} are lower than those in two-phase systems of H$_2$O+gas.\cite{yang2017molecular} This observation from MD simulations can be attributed to the presence of an excess of oil in the interface.
The presence of gases (CH$_4$/CO$_2$) in the H$_2$O-alkane interface causes a reduction of IFT with rising pressure.\cite{yang2022studyJML2,yang2023molecular,pan2023experimental,yang2005interfacial} This is in contrast to the reported pressure effect in two-phase systems of H$_2$O+alkane, where the IFT typically increases with pressure.\cite{yang2020bulk,yang2022interfacial_fuel}
Moreover, the IFTs of the CH$_4$/CO$_2$-oil interface are barely influenced by the presence of H$_2$O.
Nevertheless, there is no existing information available in the literature on the IFT of the hydrogen+water+oil three-phase mixture.

In this investigation, we carry out the first examination of the interfacial behaviors in the hydrogen+water+normal decane system in three-phase region.
To estimate the interfacial properties, we employed extensive MD simulations and density gradient theory (DGT) calculations with the Perturbed-chain Statistical Associating Fluid Theory (PC-SAFT) equation of state (EoS). The results obtained from both methods were compared for a comprehensive analysis. The results from the three-component three-phase mixture are also compared to those from the two-component two-phase mixtures to acquire a deeper understanding of the impacts of the third component.
Moreover, the impacts of temperature and pressure on interfacial properties are investigated.
The subsequent sections provide detailed information on the computation and interpretation of IFTs, density profiles, solubilities, enrichments, surface excesses, and spreading coefficients.

\section{Method}
\subsection{Molecular Simulation}
All molecular simulations in this work were conducted using the LAMMPS code.\cite{plimpton1995fast}
Our simulation system include H$_2$, H$_2$O, and n-decane (see Fig.\ref{fig:z1_vmd}).
The Mie potential is employed to describe the potential energy between segments $i$ and $j$:\cite{mie1903kinetischen}
\begin{equation}
\label{eq:mie}
U^{\mathrm{Mie}}(r_{ij}) = C_{ij} \varepsilon_{ij}\left[ \left(\frac{{\displaystyle
\sigma_{ij}}}{{\displaystyle r_{ij}}}\right)^{\lambda^r_{ij}} -
\left(\frac{{\displaystyle
\sigma_{ij}}}{{\displaystyle r_{ij}}}\right)^{\lambda^a_{ij}} \right],
\end{equation}
where $r_{ij}$ denotes the distance between segments, $\varepsilon_{ij}$ denotes the well depth, and $\sigma_{ij}$ denotes the effective segment diameter. The $\lambda^a_{ij}$ and $\lambda^r_{ij}$ are the exponents for attractive and repulsive interactions, separately. The following equation is used to compute the constant $C$:
\begin{equation}
C_{ij}=\left(\cfrac{\lambda^r_{ij}}{\lambda^r_{ij}-\lambda^a_{ij}}\right) \left(\cfrac{\lambda^r_{ij}}{\lambda^a_{ij}}\right)^{\tiny  \lambda^a_{ij}/(\lambda^r_{ij}-\lambda^a_{ij})}.
\end{equation}
The fluid components are modeled as coarse-grained (CG) segments.
While atomistic models could be devised, CG models not only accurately replicate the experimental interfacial tension,\cite{miguez2014comprehensive,garrido2019physical} but also significantly decrease the computational expenses.
The H$_2$ and H$_2$O molecules are described by single CG segments,\cite{hirsehfelder1954molecular,lobanova2015saft} while the CG model for the n-decane molecule contains three segments connected by two bonds.\cite{herdes2015coarse} The bonding energy is described using a harmonic model:
$ U_{\rm Bond}= k_{\rm bond} (r_{\rm ss}-r_{\rm 0,ss})^2$,
where $k_{\rm bond}$, $r_{\rm ss}$, and $r_{\rm 0,ss}$ are the spring constant (3333 K/\AA$^2$),\cite{rahman2018saft} distance between bonded segments, and equilibrium bond length (4.4908 \AA),\cite{lobanova2016saft} separately. The angle bending energy is also described using a harmonic model: $ U_{\rm Angle}= k_{\rm angle} (\theta-\theta_0)^2$, where $k_{\rm angle}$ is the sprint constant (1333.54 K/rad$^{-2}$),\ $\theta$ is the angle resulted from the sequential connection of three segments in one molecule, and $\theta_0$ is the equilibrium angle (157.6 \degree),\cite{herdes2018combined} respectively.
The non-bonded 1-3 interactions are included.\cite{herdes2018combined}
The force field parameters for different segments are provided in Tab. \ref{tab:ff}, Eq. \ref{eq:epsilon}, and Eq. \ref{eq:sigma}.

\begin{equation}
\label{eq:epsilon}
(\varepsilon/k_B)/\rm K = -4.806\times10^{-4}(T/K)^2 + 0.6107\times(T/K) + 165.9,
\end{equation}
\begin{equation}
\label{eq:sigma}
\sigma/\rm nm = -6.455\times10^{-10}(T/K)^3 + 9.100\times10^{-7}(T/K)^2 - 4.291\times10^{-4}(T/K)+0.3543,
\end{equation}
where $T$ is the temperature in the unit of K.
The mixing rules given by Laffitte et al.\cite{lafitte2013accurate} are employed to derive the interaction parameters between dissimilar segments:
\begin{equation}
\sigma_{ij} = \cfrac{\sigma_{ii}+\sigma_{jj}}{2},
\end{equation}
\begin{equation}
\varepsilon_{ij} = \left(1- \rm k_{ij}\right)\cfrac{\left(\sigma_{ii}^3\sigma_{jj}^3\right)^{0.5}}{\sigma_{ij}^3}\left(\varepsilon_{ii}\varepsilon_{jj}\right)^{0.5},
\end{equation}
\begin{equation}
\lambda_{ij}^l = \left[\left(\lambda_{ii}^l-3\right)\left(\lambda_{jj}^l-3\right) \right)]^{0.5} + 3   \ \  l=a, r,
\end{equation}
where $\rm k_{ij}$ refers the binary interaction parameter.
The values of $\rm k_{ij}$ for H$_2$-H$_2$O, H$_2$O-decane segment, and H$_2$-decane segment pairs are 0.26, 0.28, and 0.0, correspondingly.
The non-zero $\rm k_{ij}$ values are tuned based on experimental IFT data of 2-phase fluid mixtures.\cite{chow2018interfacial,chow2020erratum,georgiadis2011interfacial,cai1996interfacial}
A cutoff distance of 20 \AA \ was implemented for the Mie interactions among segments.

Fig. \ref{fig:z1_vmd} demonstrates equilibrium snapshots of the molecular system for studying the interfacial properties of the H$_2$+H$_2$O+decane three-phase mixture. The simulation box boundaries were set to be periodic.
The system was made of three cuboid-shaped phases, namely, the H$_2$-rich phase (connected through the periodic boundary condition), the H$_2$O-rich phase, and decane-rich phase.
There were around 2300 H$_2$O molecules, 420 decane molecules, and up to 1500 gas molecules.
The dimensions of the simulation box were set to be 44 \AA\ in both the $x$ and $y$ directions.
The size of the box in the $z$ direction was approximately 9 times greater than in the other dimensions.
These sizes have been tested to be large enough to sufficiently mitigate the size effect in this work and previous studies.\cite{yang2017molecular}
The initial positions of segments were generated using the Packmol code.\cite{martinez2009packmol}
The velocity Verlet algorithm was utilized to calculate the particle trajectories.\cite{frenkel2001understanding}
The timestep used for simulations was 5 fs.
The times for the $NPT$ equilibrium and $NVT$ production simulations were 15 ns and 75 ns, respectively.
Additionally, temperature and pressure control were separately managed using the Nos\'e-Hoover thermostat and barostat techniques.

The determination of IFTs follows the Kirkwood and Buff method,\cite{kirkwood1949statistical} which involves computing the IFT from the pressure tensor components for each interface:
\begin{equation}
\gamma = \int \Big [P_{zz}-\frac{1}{2}(P_{xx}+P_{yy})\Big ] dz.
\end{equation}
The pressure tensor components $P_{xx}$, $P_{yy}$, and $P_{zz}$ correspond to the diagonal elements.
To accurately calculate the spatial distributions of the pressure tensor using the Irving-Kirkwood contour,\cite{irving1950statistical} we utilized the package developed by Nakamura et al.\cite{nakamura2015precise} The results from production simulations were divided into 3 blocks to evaluate uncertainties.

\subsection{Density Gradient Theory}

The interfacial behaviors of the H$_2$+water+decane three-phase mixture were also investigated using DGT\cite{davis1982stress,van1894thermodynamische} with PC-SAFT EoS\cite{gross2001perturbed,gross2002application}. Details regarding DGT with PC-SAFT EoS can be found in previous studies.\cite{yang2022interfacialatmosphere,yang2020bulk}
Briefly, a three-phase flash calculation was first performed to obtain equilibrium bulk properties which serve as boundary conditions for the DGT.\cite{pan2019multiphase}
Subsequently, the DGT was employed for each pair of phases to determine the interfacial properties.\cite{mejia2021sgtpy}
Tabs. S1-S4 provide the parameters utilized in the PC-SAFT DGT.
Those parameters were taken either from previous studies\cite{alanazi2022evaluation,diamantonis2011evaluation,gross2001perturbed,yang2020bulk,yang2023molecular,mairhofer2018modeling} or fitted based on experimental data.\cite{alvarez1991semiempirical,florusse2003solubility,chow2018interfacial,chow2020erratum,georgiadis2011interfacial}
To validate our models, we compared the IFT results for two-component two-phase systems at various temperature and pressure conditions from both simulation and theory to available experiment data in the literature.\cite{chow2018interfacial,chow2020erratum,georgiadis2011interfacial,linstrom2001nist}
The IFT results obtained from MD, DGT, and experiment exhibit a good agreement as displayed in Figs. S1-S3.
Furthermore, reasonable agreement can also be seen between MD and DGT for most of the other bulk or interfacial properties including the component density profiles (Figs. S4-S6), solubilities (Figs. S7-S9), enrichments (Figs. S10-S12), and relative adsorptions (Figs. S13-S15) in the two-phase systems.

\section{Results and Discussion}

\subsection{Interface between hydrogen-rich and water-rich phases}

Initially, we investigate the H$_2$-H$_2$O interface in the three-phase mixture consisting of H$_2$, H$_2$O, and n-decane. Fig. \ref{fig:z2} illustrates the IFT values from both MD simulation and DGT for this interface. The simulation predictions align well with the results from DGT, albeit the simulation results are slightly lower than those obtained from DGT.
The IFTs values obtained from simulation and theory under the specified temperature and pressure conditions range from 55.0 to 69.6 mN/m and 50.2 to 66.7 mN/m, respectively. It is observed that the IFTs exhibit a decrease as the pressure or temperature increase, which is similar to the trends observed in prior experimental investigations of the H$_2$+H$_2$O two-phase system (also see Fig. S1).\cite{chow2018interfacial,chow2020erratum,hosseini2022h2,slowinski1957effect,massoudi1974effect}
Fig. S16(a) depicts a comparison of IFTs between the H$_2$+H$_2$O+decane three-phase mixture and the hydrogen+water two-phase mixture. Generally, the IFTs in the three-phase mixture are lower than those in the two-phase mixture. Additionally, the reductions in IFTs are more pronounced at higher temperatures within the three-phase system.
For instance, at 20 MPa, the disparity between the two IFT values computed from DGT rises from 4.9 to 5.3 mN/m when rising the temperature from 298 to 373 K.
Nevertheless, when comparing the reduction of IFTs, it is observed that the decreases in IFT values from simulations are mostly less significant compared to results obtained from the theory, particularly at lower temperatures.
Remarkably, a similar reduction trend of IFT of the gas-water interface has been reported in the previous investigations on the CH$_4$+H$_2$O+C$_{10}$H$_{22}$ and CO$_2$+H$_2$O+C$_{10}$H$_{22}$ three-phase systems.\cite{yang2022studyJML2,yang2023interfacialJML}

Due to thermal fluctuations, measuring the density distributions of components within interfacial regions poses a challenge in the experiment\cite{stephan2020enrichment}. However, in MD simulation and DGT, the spatial distributions of each component in the interface are readily accessible.\cite{stephan2020enrichment,yang2022interfacialatmosphere,yang2020bulk,yang2022interfacialJCP}.
Fig. \ref{fig:z3} illustrates the spatial distributions of hydrogen, water, and C$_{10}$H$_{22}$ obtained from both MD simulations and DGT analysis. The density profiles computed from simulations exhibit consistency with results from the theory. Moreover, the shapes of the density profiles for H$_2$ and H$_2$O resemble those observed in the H$_2$+H$_2$O two-phase mixture\cite{yang2023molecular}(see Fig. S4).

The component solubility can be evaluated based on density profiles in the bulk regions.
Figs. S17(a,b,d,e) give the solubilities in the H$_2$O-rich and hydrogen-rich phases of the three-phase mixture. The results are almost the same with solubilities in the two-phase mixtures given in Figs. S7-S9, which indicates the effects of the third component on solubilities are moderate.
Notably, the H$_2$ solubilities in the aqueous phase (Fig. S17a) from simulations are lower than data from the theory, while an opposite behavior is seen for the water solubilities in the hydrogen-rich phase (Fig. S17b). Here, the difference between MD and DGT comes mainly from the CG models used in MD simulations given the force field parameters of H$_2$O are fitted based mainly on experimental surface tension data.\cite{lobanova2015saft} Furthermore, the solubilities of decane in the aqueous phase from MD simulations given in Fig. S17d are off as expected considering the small absolute value of decane solubility.\cite{maczynski2004recommended}

As anticipated, the density profile of H$_2$O shows a continuous decrease from the aqueous phase to the hydrogen-rich phase. Conversely, a distinct peak in the spatial distribution of H$_2$ in the interfacial region is observed, particularly noticeable at 298 K. Interestingly, the interfacial region also exhibits a maximum density for C$_{10}$H$_{22}$.
Moreover, the DGT analysis predicts a greater quantity of C$_{10}$H$_{22}$ within the interfacial region. These density peaks of the oil component within the interfacial regions are associated with the boundaries of phases and are considered precursors to the emergence of a new phase.\cite{falls1983adsorption,stephan2020interfacial}
These findings align with previous investigations conducted on the 3-phase systems of the CH$_4$+water+C$_{10}$H$_{22}$ and CO$_2$+water+C$_{10}$H$_{22}$ mixtures.\cite{yang2022studyJML2,yang2023interfacialJML}

Enrichment is a commonly employed measure to evaluate the non-monotonic behavior of component density profiles:
\cite{becker2016interfacial,stephan2023monotonicity}
\begin{equation}
E_i =  \cfrac{max(n_i(z))}{max(n_i^{I},n_i^{II})},
\end{equation}
where $n_i(z)$ represents the spatial distribution of species $i$ across the interface, and $n_i^{I}$ and $n_i^{II}$ represent the densities of species $i$ in bulk phases $I$ and $II$, reparately. Fig. \ref{fig:z3e} presents the enrichments of H$_2$ and C$_{10}$H$_{22}$ in the H$_2$-H$_2$O interface. The enrichments of H$_2$ decline with pressure and temperature in line with those in the two-phase system (cf. Fig. S10).
However, the values in the three-phase mixture are generally smaller than those in the two-phase mixture, especially at low temperature and pressure conditions.
The enrichments of C$_{10}$H$_{22}$ also decline with pressure and temperature, while the corresponding values are much larger than those of H$_2$.

To comprehend the behaviors of IFT, the surface excess can be examined. The connection between the IFT and the surface excess is determined by the Gibbs adsorption equation\cite{radke2015gibbs,stephan2020enrichment,miqueu2011simultaneous,yang2019effect,y2022interfacial_JML}:
\begin{equation}
\label{eq:Gibbs}
- d\gamma = \sum\limits_{i} \Gamma_{i,j}d\mu_i,
\end{equation}
where $\Gamma_{i,j}$ denotes the surface excess of species $i$ with the reference species $j$. The expression for the surface excess is given by:\cite{telo1983structure,wadewitz1996density}
\begin{equation}
\label{eq:SurExcess}
\Gamma_{i,j} = -(n_{i}^{II} - n_{i}^{I}) \int_{-\infty}^{+\infty}  \left[ \cfrac{n_{j} (z) - n_{j}^{II}}{n_{j}^{II} - n_{j}^{I}} - \cfrac{n_{i} (z) - n_{i}^{II}}{n_{i}^{II} - n_{i}^{I}} \right] dz,\
\end{equation}
where $I$ refers to the phase enriched with component $i$ and $II$ denotes the phase enriched with component $j$.
Fig. \ref{fig:z3ra} presents the surface excesses of H$_2$ and C$_{10}$H$_{22}$ in the H$_2$-H$_2$O interface.
The surface excess of H$_2$ reduces as both pressure and temperature increase.
The same temperature effect can be seen for the surface excess of H$_2$ in the two-phase system (Fig. S13). However, the surface excess of H$_2$ in the two-phase system rises first and then declines with rising pressure. The change of pressure effects on H$_2$ surface excess comes from the presence of decane in the interfacial region which leads to less space for H$_2$.
The positive values of surface excesses of C$_{10}$H$_{22}$ account for the reduction in IFT after including the C$_{10}$H$_{22}$ into the system based on the analysis of Eq. \ref{eq:Gibbs}.
Moreover, the surface excesses of C$_{10}$H$_{22}$ are pronounced at elevated temperatures consistent with the greater difference of IFT between the two-phase and three-phase systems at higher temperatures.

\subsection{Interface between water-rich and decane-rich phases}

We then investigate the H$_2$O-C$_{10}$H$_{22}$ interface in the three-phase mixture. Fig. \ref{fig:z4} illustrates the values of IFT predicted from both MD simulation and DGT.
The simulation data corresponds closely to the predictions made by DGT.
The IFTs calculated from MD simulations and the DGT, within the specified conditions, fall within the ranges of 41.9 to 55.0 mN/m and 42.3 to 51.4 mN/m, respectively.
It is observed that the IFTs exhibit an increase as the pressure rises or temperature declines.
The pressure dependence of IFT is almost linear in the three-phase system.
Those findings are similar to the trends observed in prior experimental or simulation investigations of the H$_2$O+C$_{10}$H$_{22}$ two-phase mixture\cite{georgiadis2011interfacial,yang2020bulk} (see Fig. S2).  Remarkably, the increments of IFT with pressure from MD simulations are greater than predictions from DGT in the three-phase system, while similar slope is observed in the two-phase mixture (see Fig. S2).

Fig. S16(b) depicts a comparison of IFTs between the H$_2$+H$_2$O+decane three-phase mixture and the H$_2$O+decane two-phase mixture.
Generally, the IFTs in the three-phase mixture are lower than those in the two-phase mixture, especially at elevated pressures.
In other words, the linear slope of the IFT-pressure relation declines when H$_2$ is included in the H$_2$O-C$_{10}$H$_{22}$ interface in contrast to the H$_2$O-C$_{10}$H$_{22}$ two-phase mixture. For example, at 333 K, the linear slopes from the least-square fitting for two-phase and three-phase systems are 0.0715 (0.0536) mN/(m$\cdot$ MPa) and 0.0447 (0.0079) mN/(m$\cdot$ MPa) from MD (values in parentheses are from DGT), respectively.
A similar reduction trend of IFT of the gas-water interface has been reported in the previous investigations on the CH$_4$+H$_2$O+C$_{10}$H$_{22}$ and CO$_2$+H$_2$O+C$_{10}$H$_{22}$ three-phase systems.\cite{yang2022studyJML2,yang2023interfacialJML}
However, the reductions of IFT caused by CH$_4$ and CO$_2$ are more significant than the H$_2$ case, and increasing the pressure can even decrease the value of IFT in the water-decane interface of the three-phase mixtures with CH$_4$ or CO$_2$.\cite{yang2022studyJML2,yang2023interfacialJML}

Fig. \ref{fig:z5} illustrates the spatial distributions for the water-C$_{10}$H$_{22}$ interface. The density distributions computed from MD simulations are in line with the ones from DGT.
Moreover, the shapes of the density distributions for H$_2$O and C$_{10}$H$_{22}$ resemble those observed in the H$_2$O+C$_{10}$H$_{22}$ two-phase mixture\cite{yang2020bulk}(see Fig. S5).
The component solubility can be evaluated based on density profiles in the bulk regions.
Figs. S17(c,f) give the solubilities in the decane-rich phase of the three-phase mixture. The results are almost the same with solubilities in the two-phase systems given in Figs. S8 and S9.
Notably, the H$_2$O solubilities in the oil phase (Fig. S17f) from MD are less than values from DGT. The difference between MD and DGT could be mitigated by using an atomistic force field in the MD simulation.\cite{yang2023interfacialJML}
The density distributions of H$_2$O and C$_{10}$H$_{22}$ show a continuous change from the oil-rich phase to the aqueous phase. Notably, there is a distinct peak in the density profile of H$_2$ in the interfacial region. These findings on density distributions of H$_2$O, oil, and gas also align with previous investigations conducted on the 3-phase systems of the CH$_4$+water+decane and CO$_2$+water+decane mixtures.\cite{yang2022studyJML2,yang2023interfacialJML}

The enrichments of H$_2$ and C$_{10}$H$_{22}$ within the H$_2$O-C$_{10}$H$_{22}$ interface are depicted in Fig. \ref{fig:z5e}. The enrichments of C$_{10}$H$_{22}$ are approximately one, resembling those observed in the two-phase system (cf. Fig. S11). In addition, the enrichment values of H$_2$ are positive and decline as pressure and temperature rise.

The corresponding surface excesses of H$_2$ and C$_{10}$H$_{22}$ are illustrated in Fig. \ref{fig:z5ra}.
The surface excess of H$_2$ rises with higher pressure and decreases with higher temperatures. On the other hand, the C$_{10}$H$_{22}$ surface excess values are consistently negative compared to those in the two-phase system (Fig. S14).
However, the impact of pressure on the C$_{10}$H$_{22}$ surface excess differs between the two-phase and three-phase mixtures. In the three-phase mixture, the C$_{10}$H$_{22}$ surface excess shows a nearly linear decline with pressure, while in the two-phase mixture, it rises with pressure (cf. Fig. S14). Furthermore, the magnitude of this decrease is more pronounced at lower temperatures.
This variation in the pressure influences on the C$_{10}$H$_{22}$ surface excess is caused by the significant adsorption of H$_2$ in the interface in contrast to C$_{10}$H$_{22}$. In accordance with Equation (1), the positive values of  H$_2$ surface excess account for the reduction in IFT after adding H$_2$ into the H$_2$O-C$_{10}$H$_{22}$ interface.

\subsection{Interface between hydrogen-rich and decane-rich phases}

Fig. \ref{fig:z6} depicts the IFT values calculated from both molecular simulations and DGT for the H$_2$-C$_{10}$H$_{22}$ interface. The IFT values exhibit a resemblance to those observed in the hydrogen+decane two-phase mixture (cf. Fig. S3). Specifically, the differences in IFT values from the MD simulations are moderate. Moreover, the disparities in IFT values from the DGT method are also minor, considering that the IFT curves for the three-phase mixture and the two-phase mixture overlap with each other, as demonstrated in Fig. S16(c).
Fig. \ref{fig:z7} illustrates the density distributions of each component within the three-phase systems. Upon comparison with the spatial distributions in the H$_2$+C$_{10}$H$_{22}$ two-phase mixture shown in Fig. S6, only minor differences are noticeable.
The densities of H$_2$O are relatively low in contrast to the densities of the other components. Consequently, the limited influence of H$_2$O on IFT can be attributed to its low solubility in both the decane-rich phase\cite{maczynski2004recommended} and the H$_2$-rich phase\cite{eller2022modeling,yang2023molecular} under relatively high pressure conditions.
The enrichments of H$_2$ in the three-phase systems given in Fig. \ref{fig:z7e}a exhibit nearly identical trends to those observed in the two-phase mixture in Fig. S12. Interestingly, H$_2$O enrichments are relatively large and decrease as pressure and temperature rise given in Fig. \ref{fig:z7e}b. Additionally, the disparities between the H$_2$ surface excesses and those of C$_{10}$H$_{22}$ (the reference species) in the three-phase mixture (Fig. \ref{fig:z7ra}a) and the two-phase mixture (Fig. S15) are also negligible. It is worth noting that H$_2$O surface excesses displayed in Fig. \ref{fig:z7ra}b have positive values, and exhibit an increment with rising temperature.

\subsection{Spreading Coefficient}

The spreading coefficient, denoted as $S$, can be employed to determine whether the oil has the ability to form a thin film that spreads between phases.\cite{shaw1980introduction,harkins1922films}
The estimation of $S$ can be accomplished by utilizing the IFTs obtained from the three-phase mixtures:
\begin{equation}
S = \lambda_{wg}-\lambda_{wo}-\lambda_{og},
\label{eq:SS}
\end{equation}
where $\lambda_{wg}$ represents the IFT of the water-gas interface, $\lambda_{wo}$ represents the IFT of the oil-water interface, and $\lambda_{og}$ represents the IFT of the gas-oil interface of the three-phase mixture.
When a positive value of $S$ is obtained, it indicates the ability of the oil film to spread and create a separation between the H$_2$-rich phase and the H$_2$O-rich phase. For our simulation system (see Fig. \ref{fig:z1_vmd}), a positive $S$ may lead to the formation of a new oil phase separating the water-rich and gas-rich phases because of the lower interfacial free energy.
Conversely, if the value of $S$ is negative, it indicates the formation of an oil droplet.\cite{bonn2009wetting,bertrand2001wetting}

Fig. \ref{fig:z8} displays the values of $S$ in the hydrogen+water+C$_{10}$H$_{22}$ 3-phase mixture.
It is found that values of $S$ grow with rising pressure and temperature.
The values of $S$ obtained from MD are consistent with data predicted by DGT, and fall in ranges from -3.5 to 1.0 mN/m and from -7.4 to -1.5 mN/m for MD and DGT, separately.
The values of $S$ from MD are systematically larger than those from DGT because of larger $\lambda_{wg}$ values (see Fig. \ref{fig:z2}).
The majority of the $S$ values are negative, which suggests the occurrence of oil droplet.
At 373 K and around 100 MPa, the $S$ is at around 1.0 mN/m. However, the new oil phase is not observed by checking the density profiles, possibly due to the relatively moderate value of $S$. In this case, our system corresponds to a metastable three-phase state.

\section{Conclusion}

This article investigates the interfacial characteristics of the H$_2$+H$_2$O+n-decane three-phase mixture under various temperature (298, 333, and 373 K) and pressure (up to approximately 110 MPa) conditions.
To estimate the interfacial properties, both MD simulation and DGT modeling using the PC-SAFT equation of state are employed. Remarkably, these two methods yield consistent estimations, demonstrating good agreement between them.
Our findings reveal that IFTs of the gas-water interface in the hydrogen+water+n-decane three-phase mixture are lower compared to IFTs in the hydrogen+water two-phase mixture. This decrease in IFT can be attributed to the excess of the oil component in the interfacial region.
In addition, hydrogen accumulates in the interfacial region of the water-oil interface in the three-phase mixture, which weakens the increment of IFT with increasing pressure compared to IFTs in the H$_2$O+oil two-phase mixture.
The IFTs of the hydrogen-oil interface are barely influenced by H$_2$O due to the limited amount of dissolved H$_2$O in bulk phases. Nevertheless, relatively strong enrichments and positive surface excesses of water are observed in the hydrogen-oil interfacial region.
Moreover, the spreading coefficients predominantly exhibit negative values, which suggests the existence of the three-phase contact for the hydrogen+water+n-decane mixture.
This research provides significant insights into the fundamental understanding of interfacial phenomena in the three-phase systems involving hydrogen, water, and oil. These findings can be utilized to enhance the design of processes related to the underground storage of hydrogen in depleted oil fields.


\bigskip
{\bf{Acknowledgments\\[1ex]}}
This project is supported by the National Natural Science Foundation of China under Grant No. 42203041, the Natural Science Foundation of Jiangsu Province under Grant No. BK20221132, and the China Postdoctoral Science Foundation under Grant No. 2022M723398.

{\bf{Supporting Information\\[1ex]}}
Additional tables and figures are given in the Supplementary Materials.

\bibliography{New}

\clearpage
{\renewcommand{\arraystretch}{0.7}
\begin{threeparttable}
\centering\caption{Mie force field parameters of different fluid components.}
\label{tab:ff}
{\begin{tabular}{@{}cccccc}
\toprule
Segment & $\varepsilon_{ii}/k_B$ (K) & $\sigma_{ii}$ (nm) & $\lambda^r_{ii}$ & $\lambda^a_{ii}$ & Reference\\
\bottomrule
H$_2$ & 33.3133 &  0.29700 &  12.0 & 6.0 & \cite{hirsehfelder1954molecular}\\
H$_2$O & Eq. \ref{eq:epsilon} & Eq. \ref{eq:sigma} &  8.0 & 6.0 & \cite{lobanova2015saft}\\
C$_{10}$H$_{22}$ & 415.1900 & 0.45841 &  20.9 & 6.0 & \cite{herdes2015coarse}
\\
\bottomrule
\end{tabular}}
\end{threeparttable}

\clearpage
\begin{figure}[tb]
\begin{centering}
\includegraphics[width=0.95\textwidth]{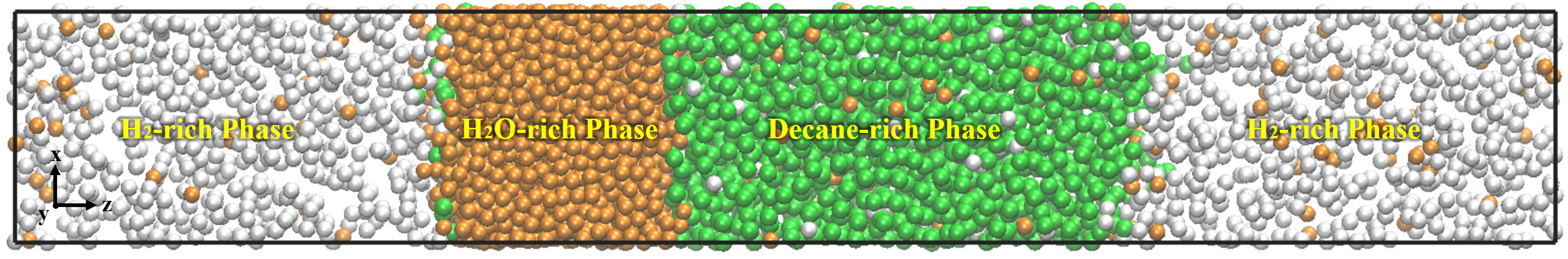}
\caption{Equilibrium snapshots of the H$_2$+H$_2$O+n-decane three-phase system  at 333 K and 20 MPa. Color code: H$_2$, white; H$_2$O, orange; n-decane segment, green.
}
\label{fig:z1_vmd}
\end{centering}
\end{figure}

\clearpage
\begin{figure}[tb]
\begin{centering}
\includegraphics[width=0.8\textwidth]{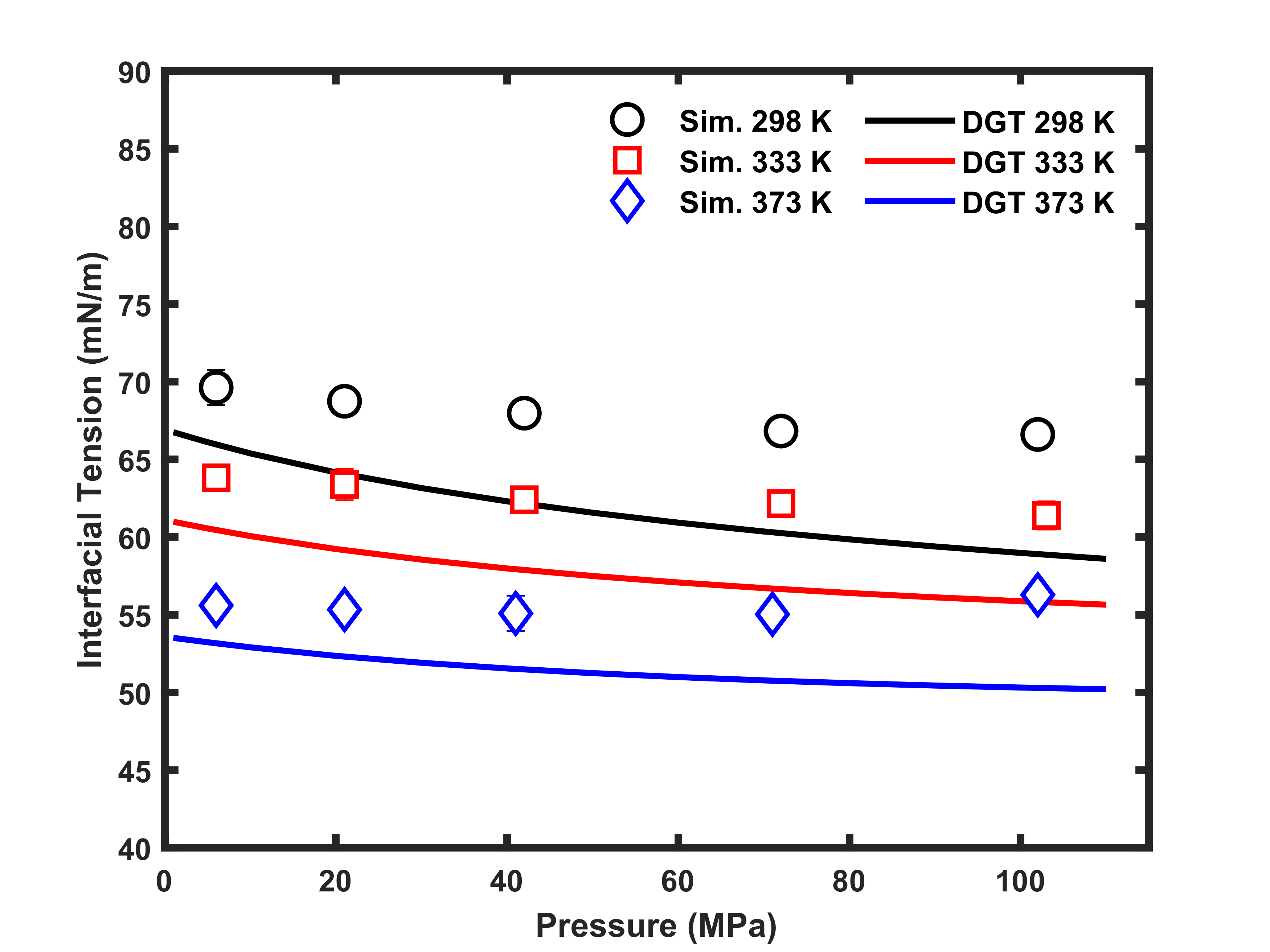}
\caption{Interfacial tensions as a function of pressure in the H$_2$+H$_2$O+C$_{10}$H$_{22}$ 3-phase system for the interface between H$_2$O-rich phase and H$_2$-rich phase. The open symbols represent the data from the MD simulation, and the data obtained using DGT with the PC-SAFT EoS are shown as lines. Error bars smaller than the symbol size are not displayed.}
\label{fig:z2}
\end{centering}
\end{figure}

\clearpage
\begin{figure}[tb]
\begin{centering}
\includegraphics[width=1.0\textwidth]{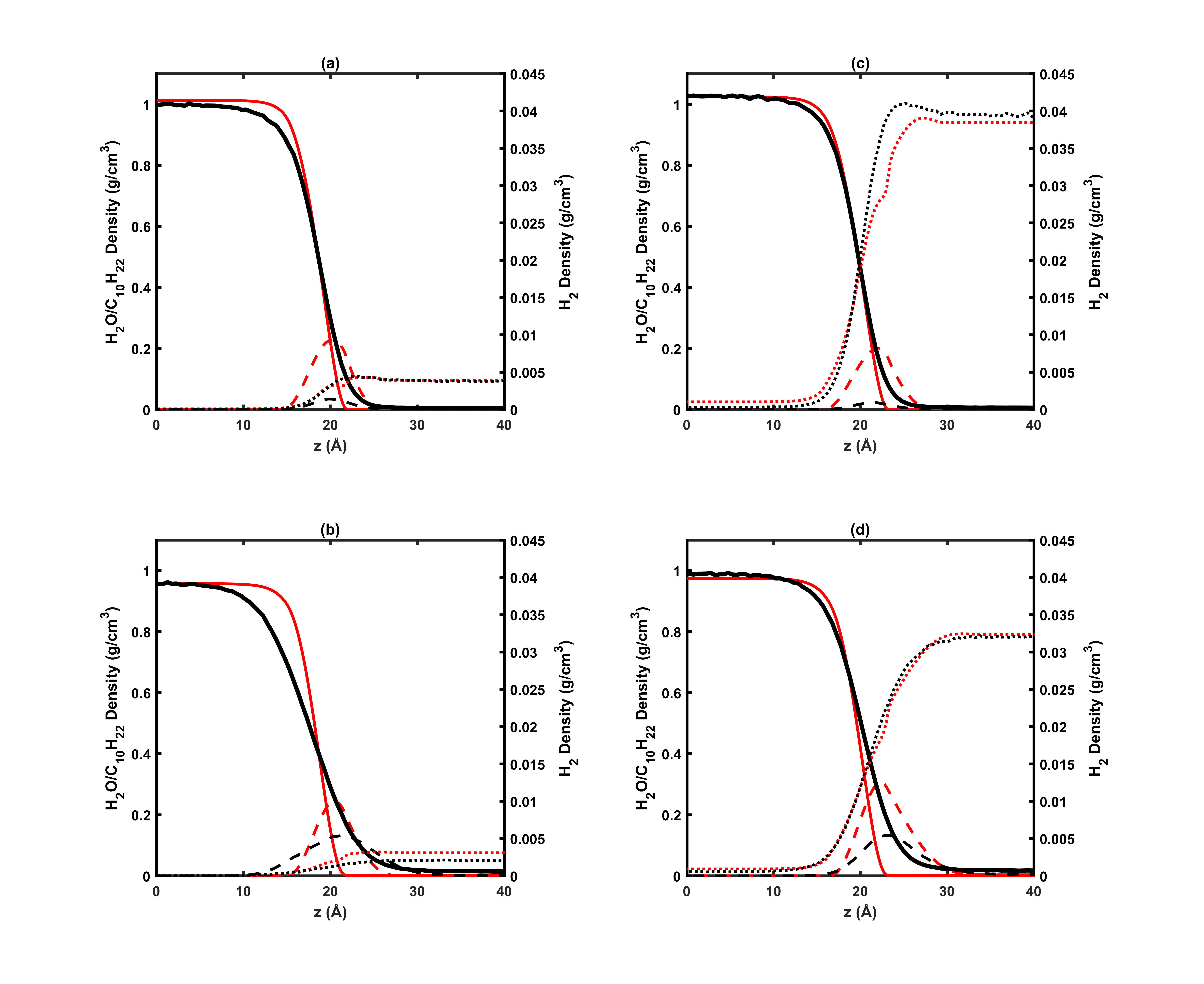}
\caption{Equilibrium distributions of different species in the   H$_2$+H$_2$O+C$_{10}$H$_{22}$ 3-phase system for the interface between H$_2$O-rich phase and vapor phase at (a) 298 K, 5 MPa, (b) 373 K, 5 MPa, (c) 298 K, 70 MPa, and (d) 373 K, 70 MPa.
The black and red colors denote MD and DGT data, respectively. The solid, dotted, and dashed lines represent H$_2$O, H$_2$, and C$_{10}$H$_{22}$, respectively.}
\label{fig:z3}
\end{centering}
\end{figure}

\clearpage
\begin{figure}[tb]
\begin{centering}
\includegraphics[width=0.5\textwidth]{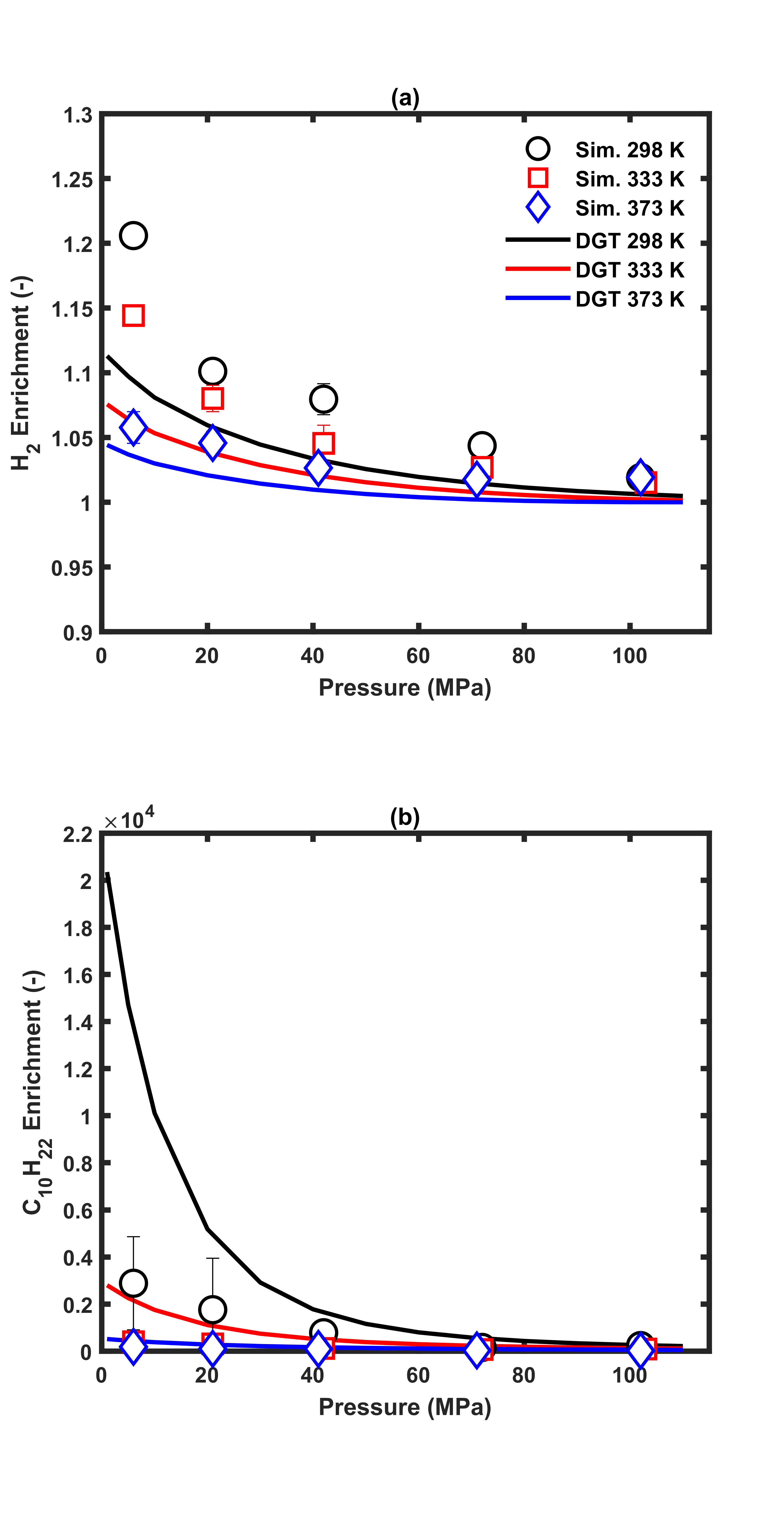}
\caption{Pressure dependence of component enrichment for the interface between H$_2$-rich phase and H$_2$O-rich phase in the H$_2$+H$_2$O+C$_{10}$H$_{22}$ 3-phase system.
The open symbols represent the data from the MD simulation, and the data from DGT with the PC-SAFT EoS are shown as lines.}
\label{fig:z3e}
\end{centering}
\end{figure}

\clearpage
\begin{figure}[tb]
\begin{centering}
\includegraphics[width=0.5\textwidth]{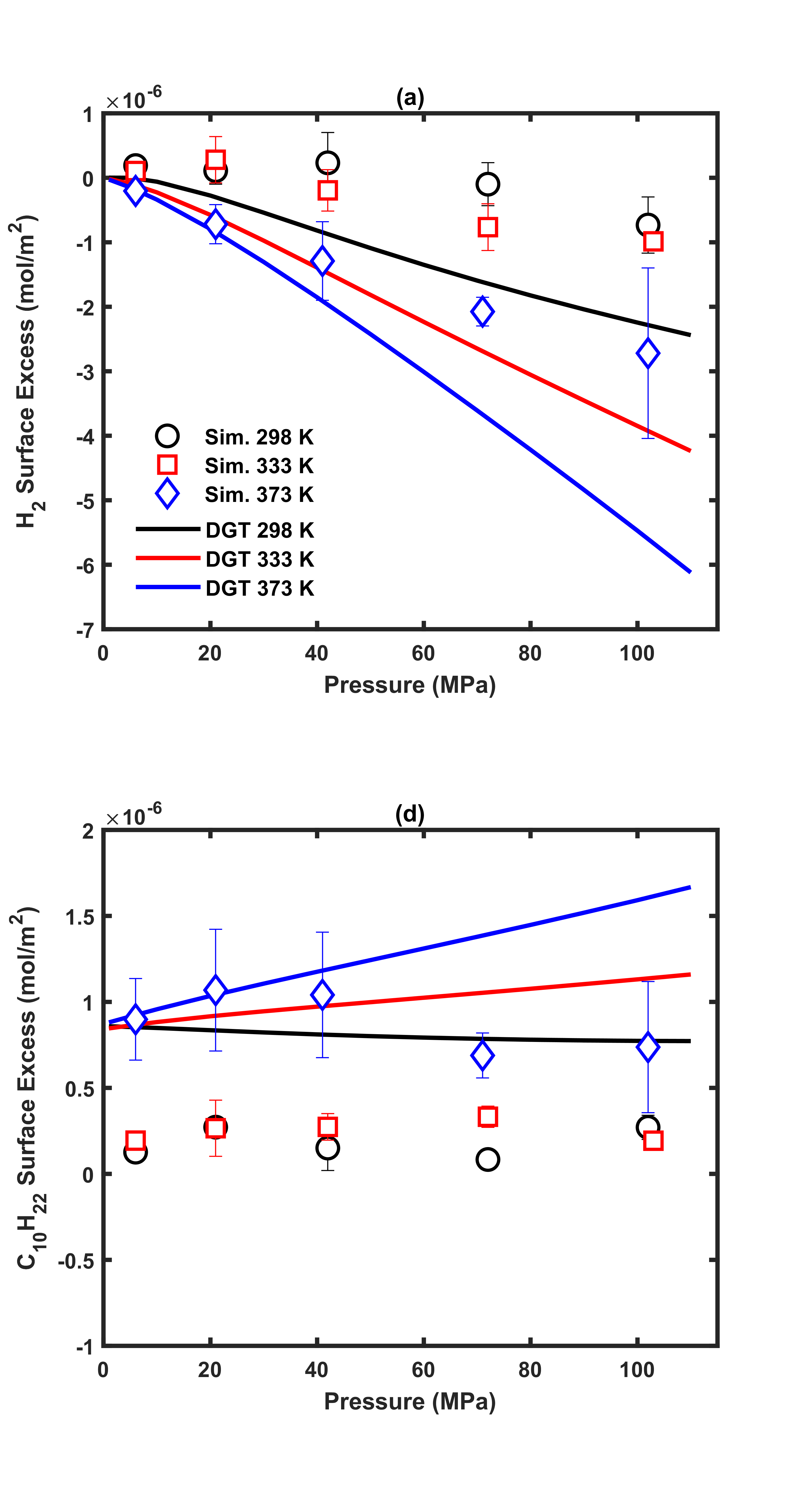}
\caption{Pressure dependence of component surface excess for the interface between H$_2$-rich phase and H$_2$O-rich phase in the H$_2$+H$_2$O+C$_{10}$H$_{22}$ 3-phase system.
The open symbols represent the data from the MD simulation, and the data from DGT with the PC-SAFT EoS are shown as lines.}
\label{fig:z3ra}
\end{centering}
\end{figure}

\clearpage
\begin{figure}[tb]
\begin{centering}
\includegraphics[width=0.8\textwidth]{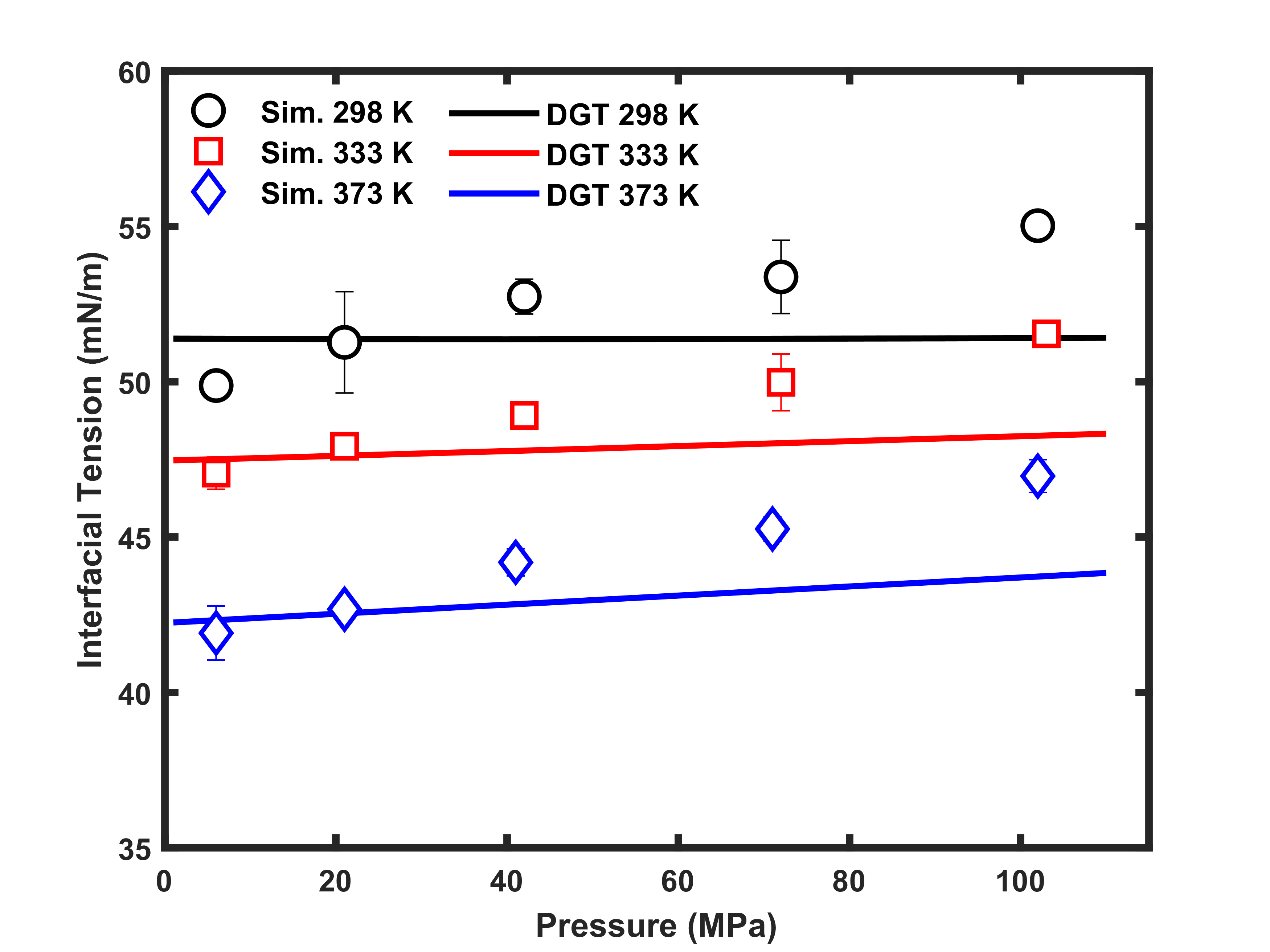}
\caption{Interfacial tensions as a function of pressure in the H$_2$+H$_2$O+C$_{10}$H$_{22}$ 3-phase system for the interface between H$_2$O-rich phase and C$_{10}$H$_{22}$-rich phase. The open symbols represent the data from the MD simulation, and the data obtained using DGT with the PC-SAFT EoS are shown as lines. Error bars smaller than the symbol size are not displayed.
}
\label{fig:z4}
\end{centering}
\end{figure}

\newpage
\begin{figure}[tb]
\begin{centering}
\includegraphics[width=1.0\textwidth]{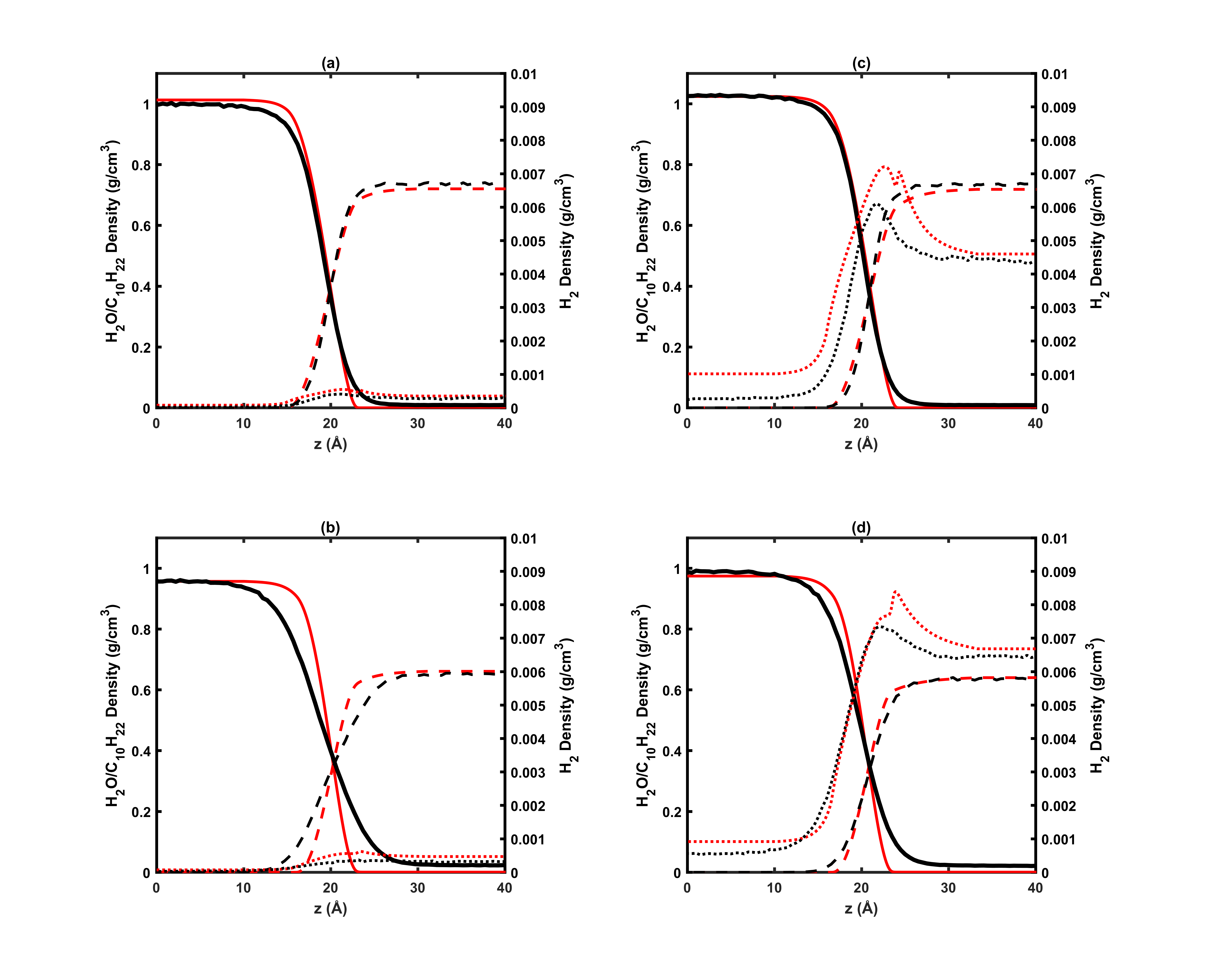}
\caption{Equilibrium distributions of different species in the   H$_2$+H$_2$O+C$_{10}$H$_{22}$ 3-phase system for the interface between H$_2$O-rich phase and C$_{10}$H$_{22}$-rich phase at (a) 298 K, 5 MPa, (b) 373 K, 5 MPa, (c) 298 K, 70 MPa, and (d) 373 K, 70 MPa.
The black and red colors denote MD and DGT data, respectively. The solid, dotted, and dashed lines represent H$_2$O, H$_2$, and C$_{10}$H$_{22}$, respectively
}
\label{fig:z5}
\end{centering}
\end{figure}

\clearpage
\begin{figure}[tb]
\begin{centering}
\includegraphics[width=0.5\textwidth]{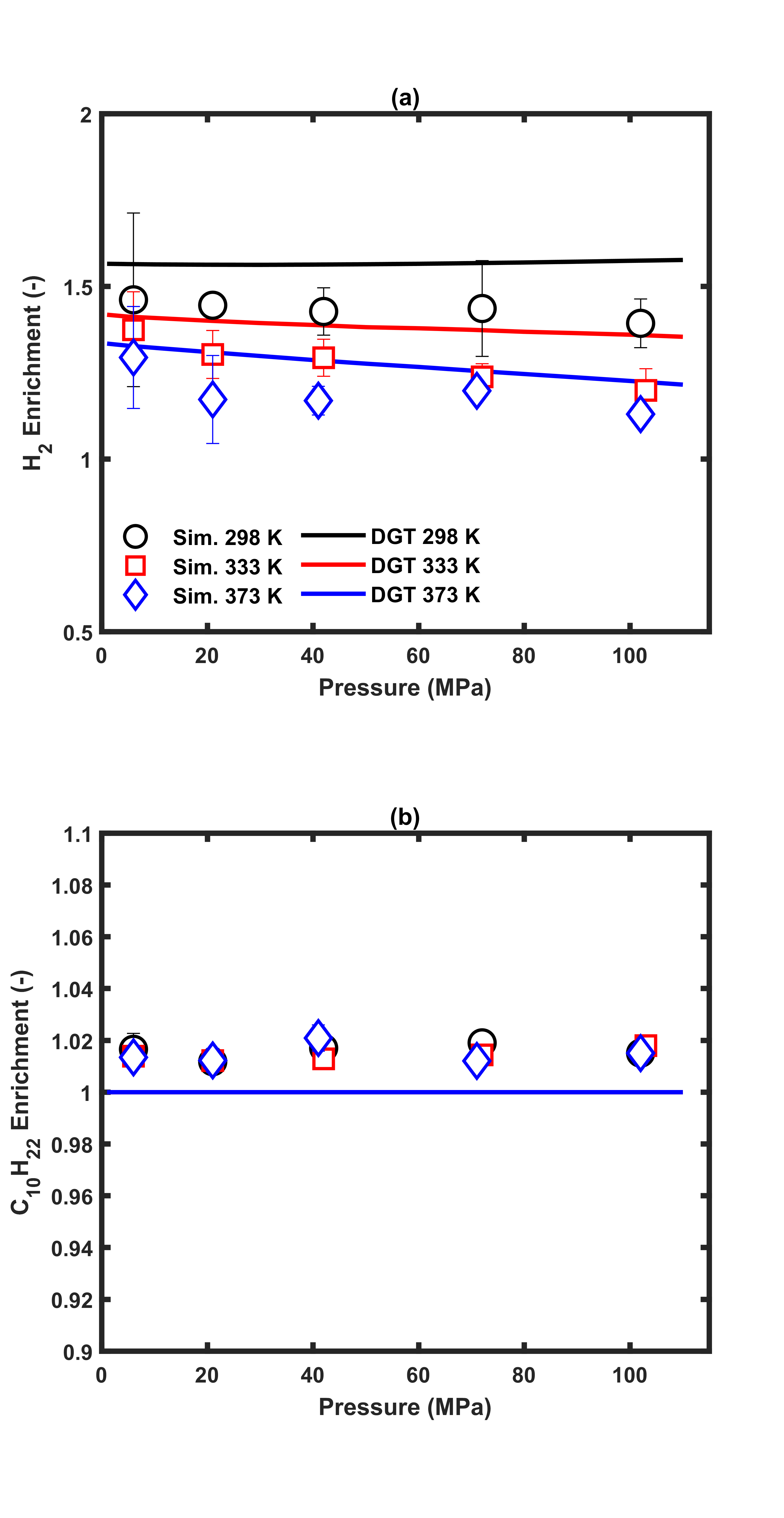}
\caption{Pressure dependence of component enrichment for the interface between H$_2$O-rich phase and C$_{10}$H$_{22}$-rich phase in the H$_2$+H$_2$O+C$_{10}$H$_{22}$ 3-phase system.
The open symbols represent the data from the MD simulation, and the data from DGT with the PC-SAFT EoS are shown as lines.}
\label{fig:z5e}
\end{centering}
\end{figure}

\clearpage
\begin{figure}[tb]
\begin{centering}
\includegraphics[width=0.5\textwidth]{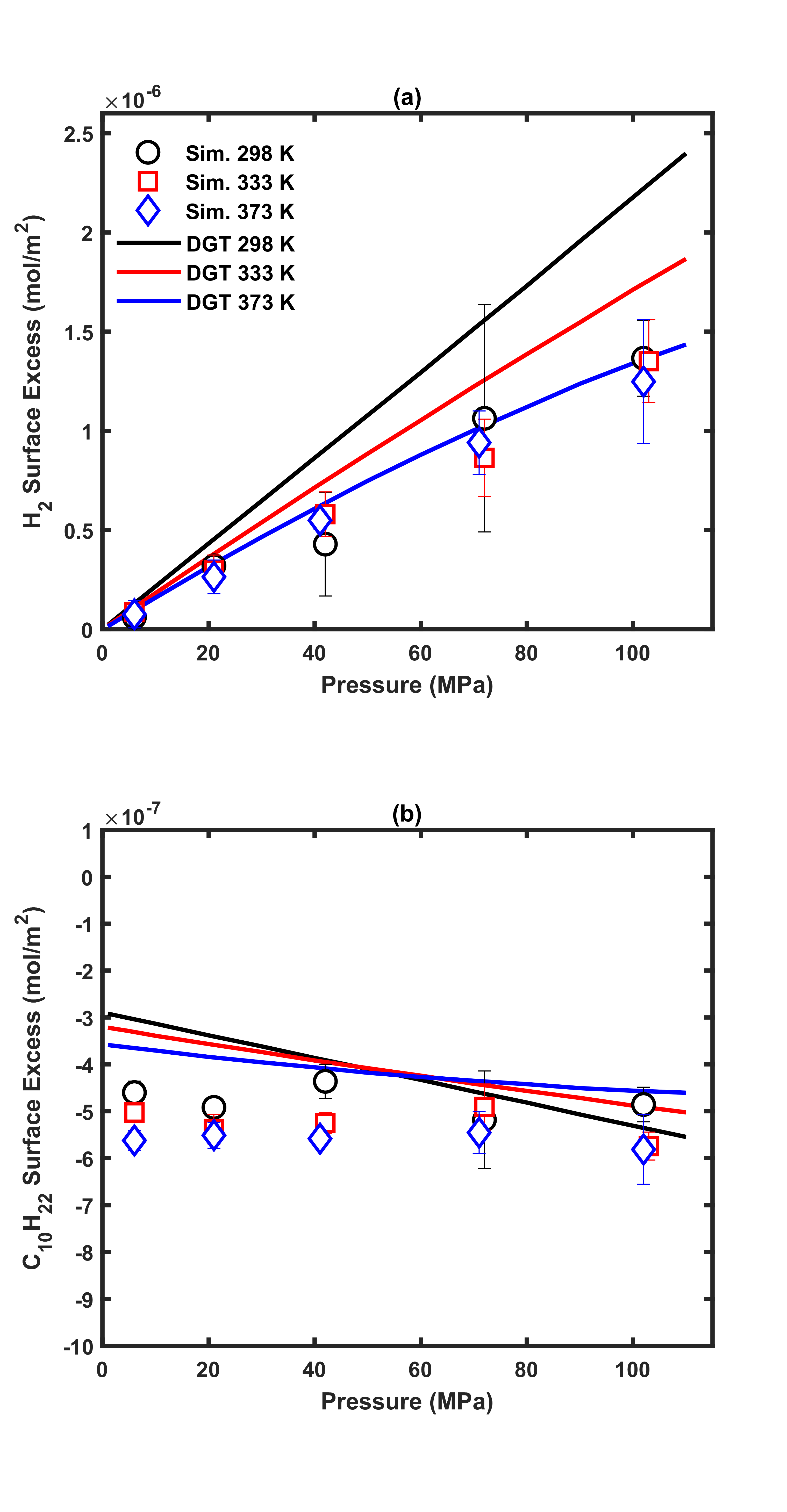}
\caption{Pressure dependence of component surface excess for the interface between H$_2$O-rich phase and C$_{10}$H$_{22}$-rich phase in the H$_2$+H$_2$O+C$_{10}$H$_{22}$ 3-phase system.
The open symbols represent the data from the MD simulation, and the data from DGT with the PC-SAFT EoS are shown as lines.}
\label{fig:z5ra}
\end{centering}
\end{figure}

\newpage
\begin{figure}[tb]
\begin{centering}
\includegraphics[width=0.8\textwidth]{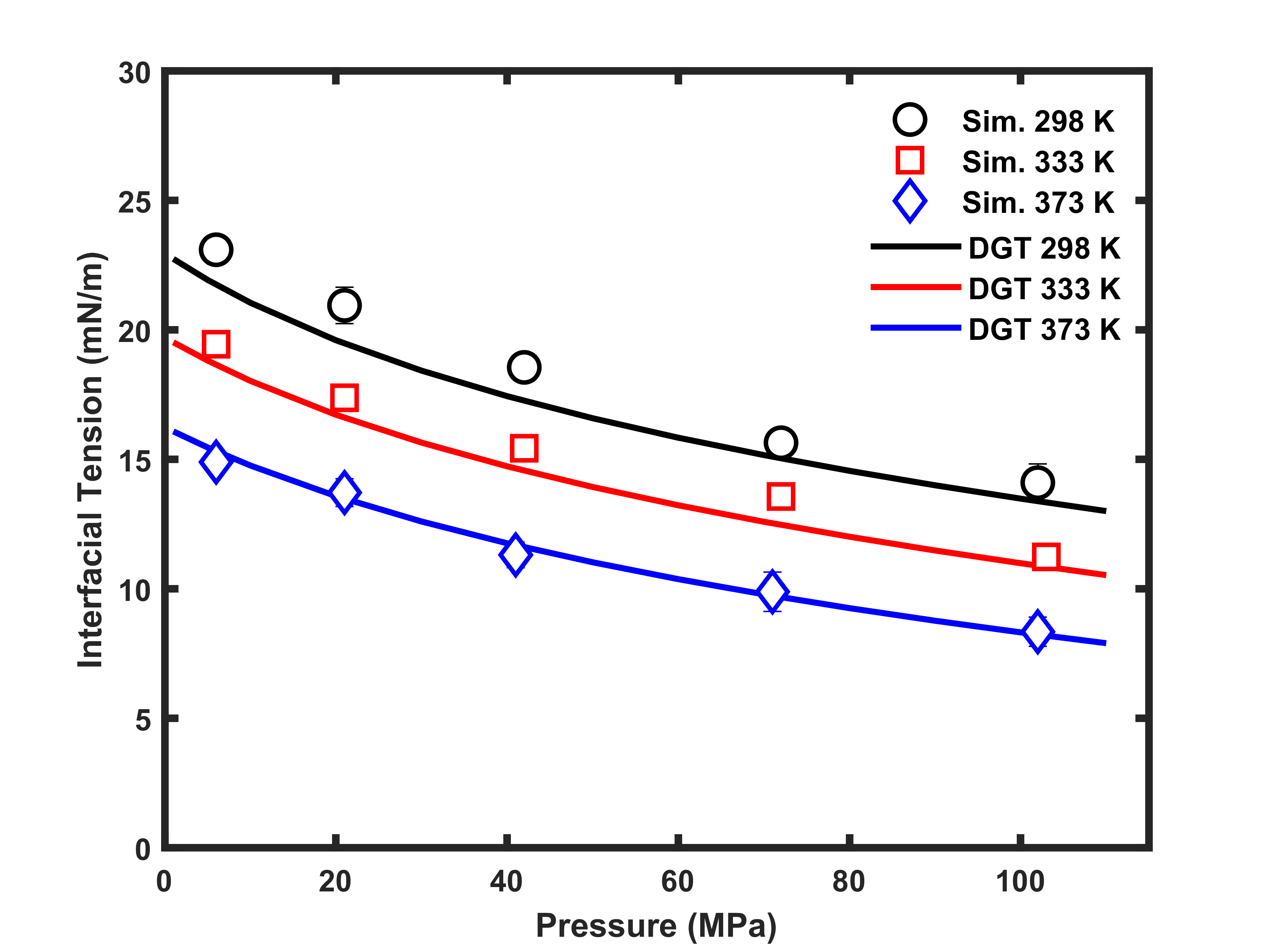}
\caption{Interfacial tensions as a function of pressure in the H$_2$+H$_2$O+C$_{10}$H$_{22}$ 3-phase system for the interface between H$_2$-rich phase and C$_{10}$H$_{22}$-rich phase. The open symbols represent the data from the MD simulation, and the data obtained using DGT with the PC-SAFT EoS are shown as lines. Error bars smaller than the symbol size are not displayed.
}
\label{fig:z6}
\end{centering}
\end{figure}

\newpage
\begin{figure}[tb]
\begin{centering}
\includegraphics[width=1.0\textwidth]{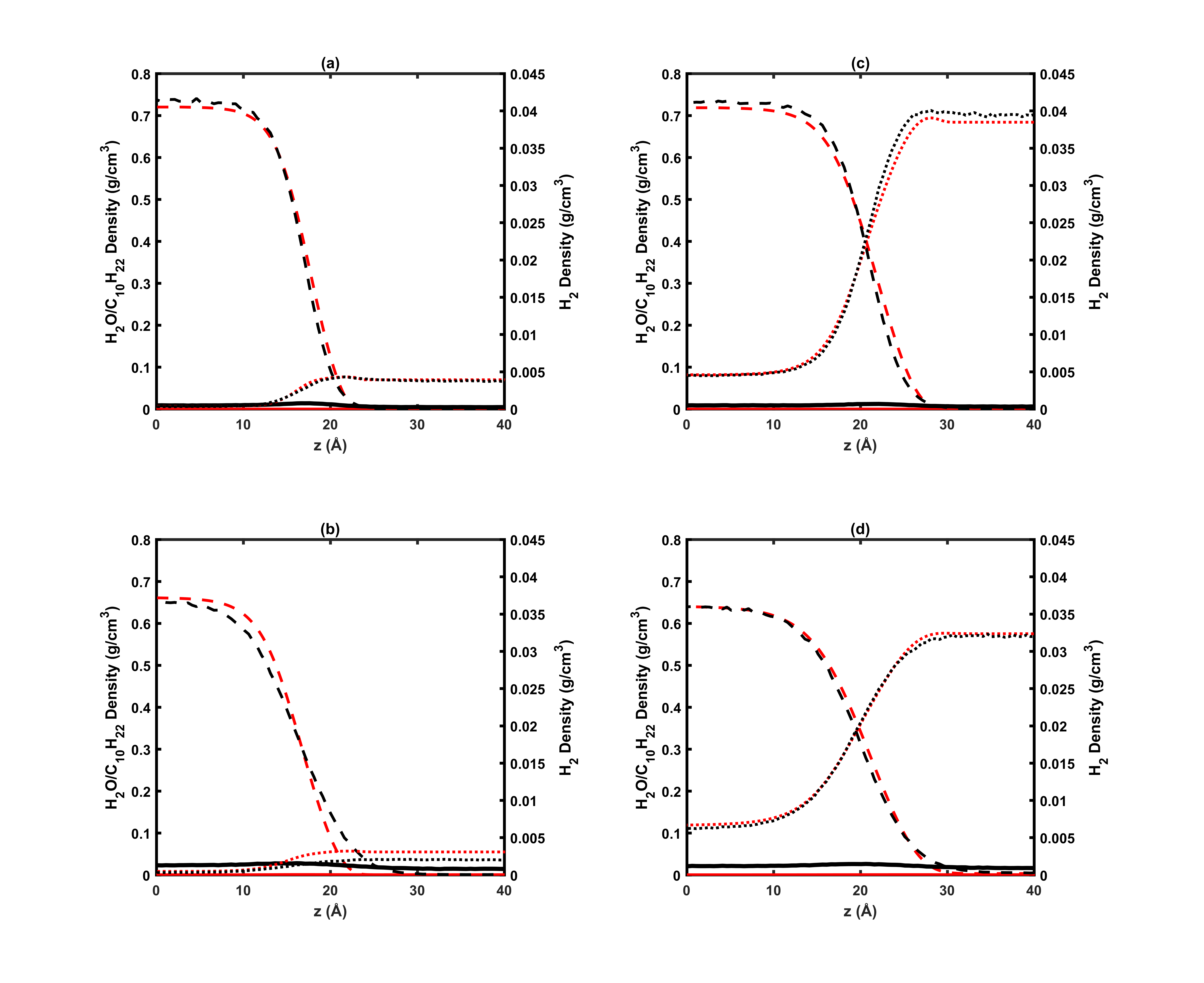}
\caption{Equilibrium distributions of different species in the   H$_2$+H$_2$O+C$_{10}$H$_{22}$ 3-phase system for the interface between H$_2$-rich phase and C$_{10}$H$_{22}$-rich phase at (a) 298 K, 5 MPa, (b) 373 K, 5 MPa, (c) 298 K, 70 MPa, and (d) 373 K, 70 MPa.
The black and red colors denote MD and DGT data, respectively. The solid, dotted, and dashed lines represent H$_2$O, H$_2$, and C$_{10}$H$_{22}$, respectively
}
\label{fig:z7}
\end{centering}
\end{figure}

\clearpage
\begin{figure}[tb]
\begin{centering}
\includegraphics[width=0.5\textwidth]{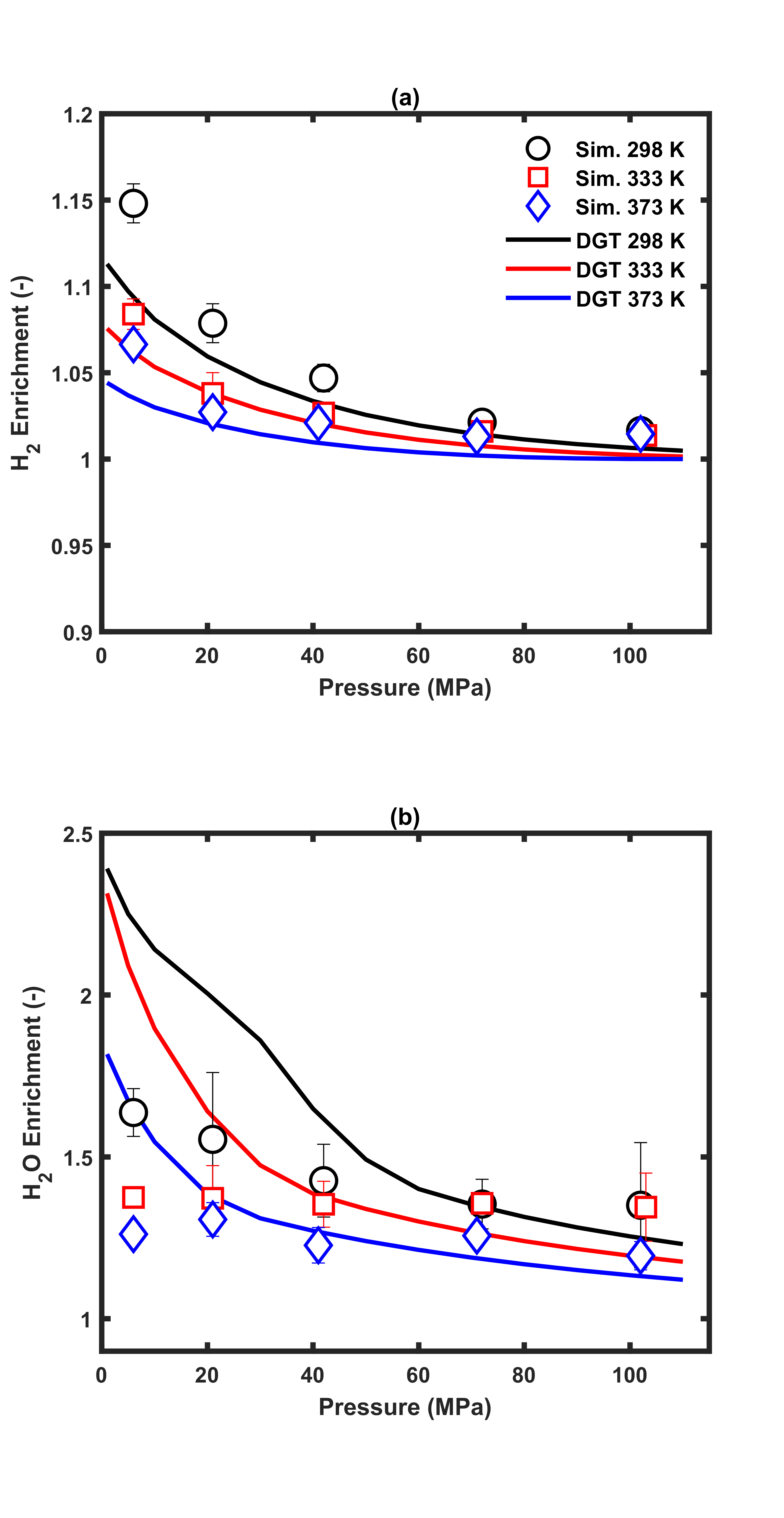}
\caption{Pressure dependence of component enrichment for the interface between H$_2$-rich phase and C$_{10}$H$_{22}$-rich phase in the H$_2$+H$_2$O+C$_{10}$H$_{22}$ 3-phase system.
The open symbols represent the data from the MD simulation, and the data from DGT with the PC-SAFT EoS are shown as lines.}
\label{fig:z7e}
\end{centering}
\end{figure}

\clearpage
\begin{figure}[tb]
\begin{centering}
\includegraphics[width=0.5\textwidth]{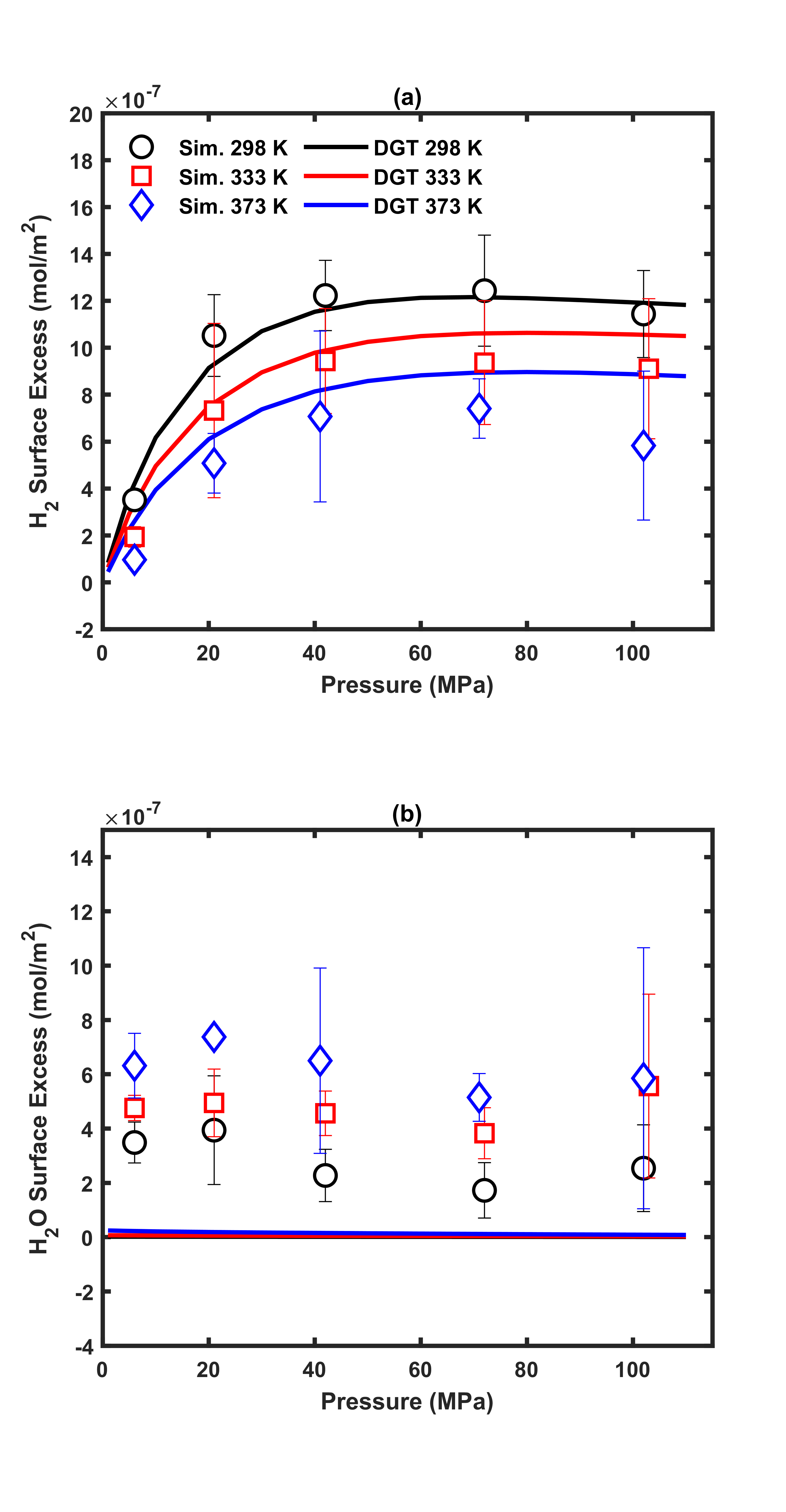}
\caption{Pressure dependence of component surface excess for the interface between H$_2$-rich phase and C$_{10}$H$_{22}$-rich phase in the H$_2$+H$_2$O+C$_{10}$H$_{22}$ 3-phase system.
The open symbols represent the data from the MD simulation, and the data from DGT with the PC-SAFT EoS are shown as lines.}
\label{fig:z7ra}
\end{centering}
\end{figure}

\newpage
\begin{figure}[tb]
\begin{centering}
\includegraphics[width=0.8\textwidth]{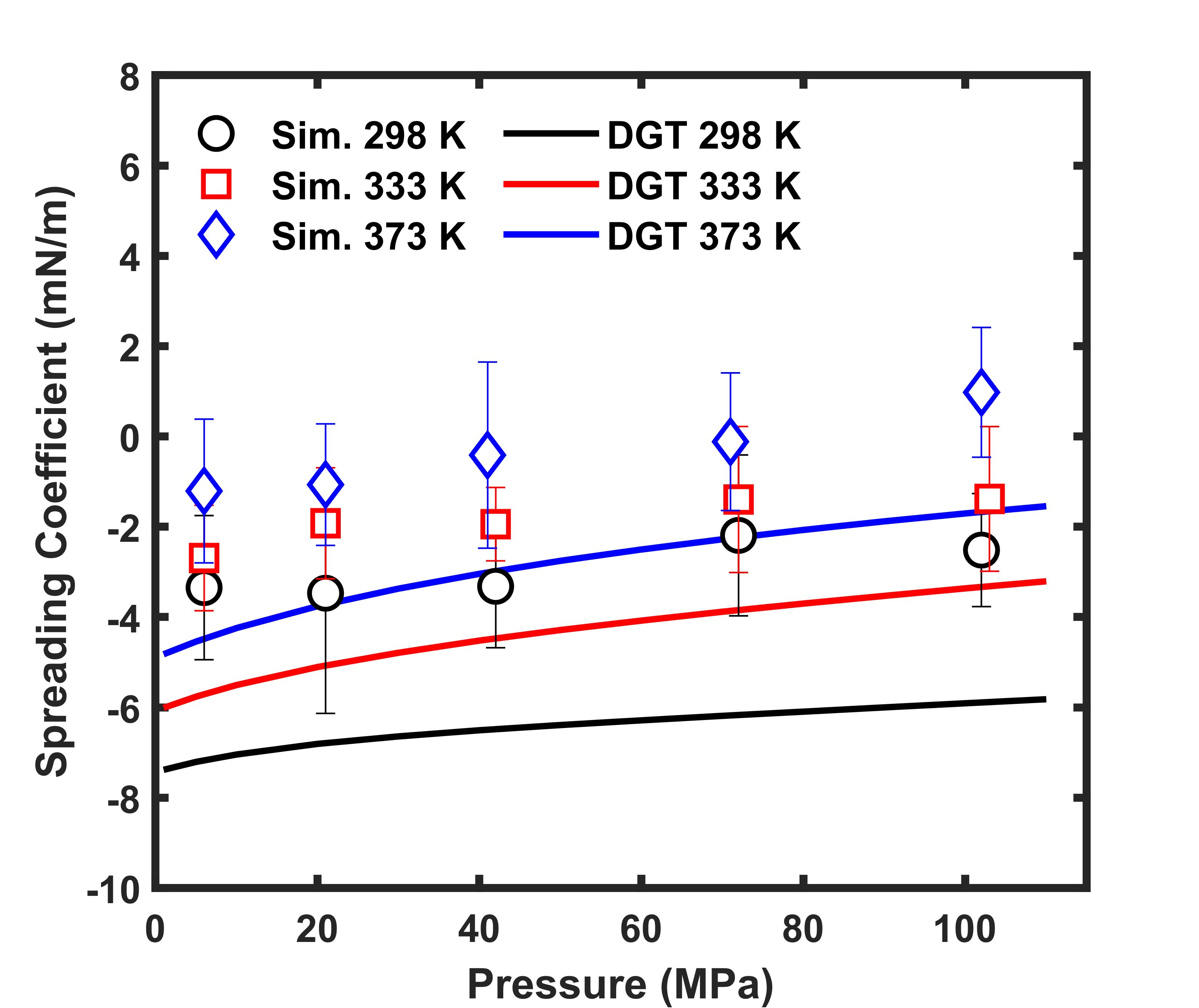}
\caption{Spreading coefficient as a function of pressure in the H$_2$+H$_2$O+C$_{10}$H$_{22}$ 3-phase system. The open symbols represent the results from the MD simulation, and the estimates obtained using DGT with the PC-SAFT EoS are shown as lines.
}
\label{fig:z8}
\end{centering}
\end{figure}

\newpage
\begin{figure}[tb]
\begin{centering}
\centerline{\includegraphics[width=1.1\textwidth]{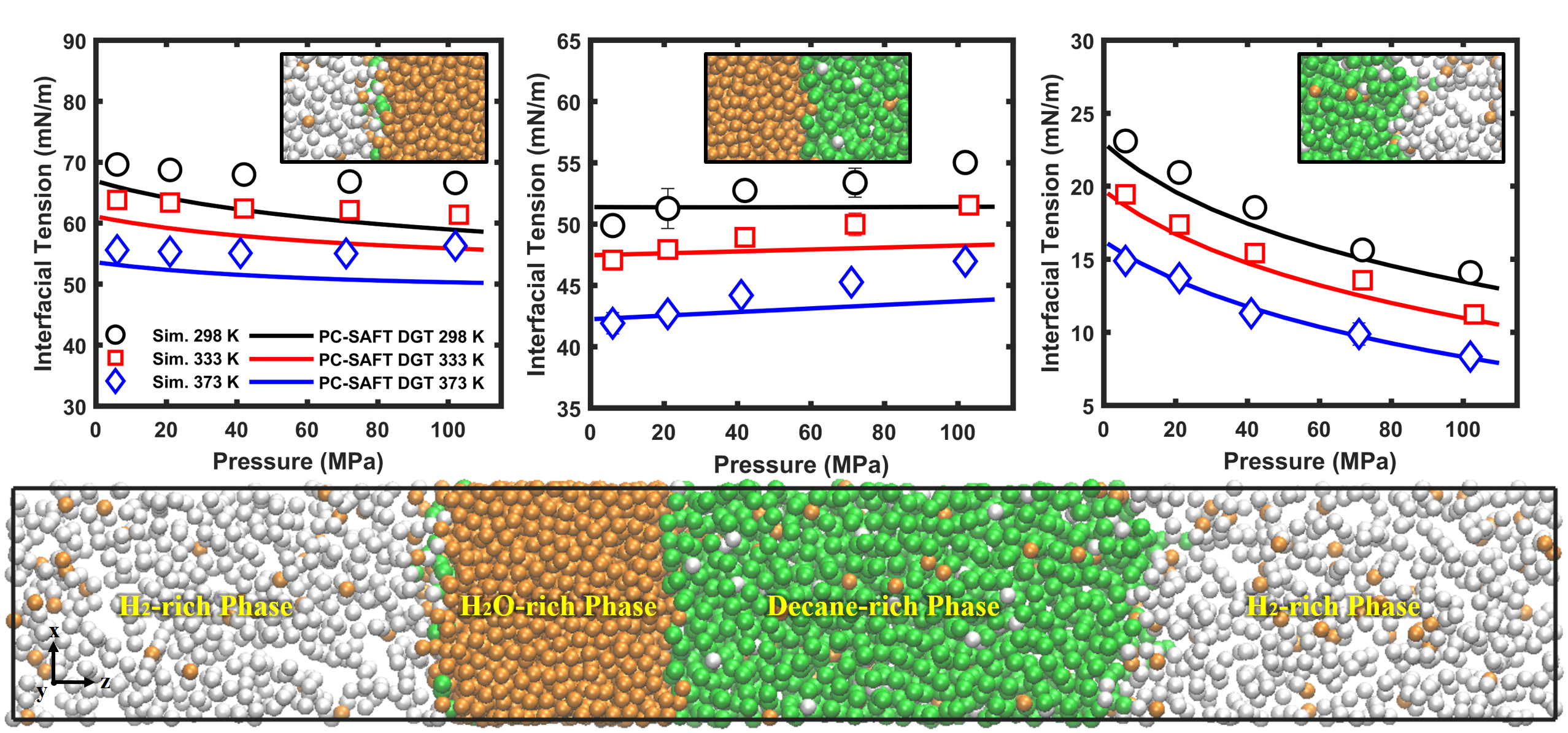}}
\vskip0.5cm
\end{centering}
\vskip3cm
{\large \bf TOC Graphic\\[2ex]}
\end{figure}

\end{document}